\RequirePackage{ifpdf}
\ifpdf 
\documentclass[pdftex]{sigma}
\else
\documentclass{sigma}
\fi

\begin{document}
\allowdisplaybreaks

\renewcommand{\PaperNumber}{024}

\def\d{\mbox{\rm d}}

\FirstPageHeading

\ShortArticleName{Symmetry Properties  of Autonomous Integrating Factors}

\ArticleName{Symmetry Properties\\ of Autonomous Integrating Factors}

\Author{Sibusiso MOYO~$^\dag$ and P.G.L. LEACH~$^\ddag$}
\AuthorNameForHeading{S. Moyo and P.G.L. Leach}

\Address{$^\dag$~Department
of Mathematics, Durban Institute of Technology,\\
$\phantom{^\dag}$~PO Box 953, Steve Biko Campus, Durban 4000, Republic of South Africa}
\EmailD{\href{mailto:moyos@dit.ac.za}{moyos@dit.ac.za}}
\Address{$^\ddag$~School of Mathematical Sciences, Howard College,
University of KwaZulu-Natal,\\
$\phantom{^\ddag}$~Durban 4041, Republic of South Africa}
\EmailD{\href{mailto:leachp@ukzn.ac.za}{leachp@ukzn.ac.za}}

\ArticleDates{Received September 27, 2005, in final form November 21,
2005; Published online December 05, 2005}

\Abstract{We study the symmetry properties of
autonomous integrating factors from an algebraic point of view.
The symmetries are delineated for the resulting integrals treated
as equations and symmetries of the integrals treated as functions
or configurational invariants. The succession of terms (pattern)
is noted. The general pattern for the solution symmetries for
equations in the simplest form of maximal order is given and the
properties of the associated integrals resulting from this
analysis are given.}


\Classification{34A05; 34A30; 34C14; 34C20} 

\section{Introduction}
It is well-known that, when a symmetry is used to determine a
first integral for a differential equation, the symmetry provides
an integrating factor for the equation and remains as a~sym\-met\-ry
of the first integral.
For first-order ordinary differential
equations the direct determination of the integrating factor is
known \cite{blumankumei} and algorithms for finding integrating factors for equations
of higher order have been developed. In 1999 Cheb-Terrab and
Roche \cite{chebroche} presented a syste\-ma\-tic algorithm for the
construction of integrating factors for second-order ordinary
differential equations and claimed that their algorithm gave
integrating factors for equations which did not possess Lie point
symmetries. In 2002 Leach and Bouquet \cite{leachbouquet} showed
that for all equations except one of which Cheb-Terrab and Roche \cite{chebroche} 
had found integrating factors had symmetries which were not
necessarily point symmetries but generalised or nonlocal. In
the same year, 2002, Abraham-Shrauner \cite{abraham} also wrote a paper to
demonstrate the reduction of order of nonlinear ordinary
differential equations by a combination of first integrals and Lie
group symmetries. The latter and former motivated us hereby to
investigate the underlying
properties of autonomous integrating factors and the associated
integrals treated as equations and as functions.\  Observations
are made and inferred in general for any $n$th-order ordinary
differential equation of maximal symmetry. These will also be
extended to include other types of equations in a separate contribution.

Program {\tt LIE} \cite{head} is used to compute the symmetries  
for the different cases considered. The knowledge of the 
symmetries of first integrals of the equation does give rise to 
some interesting properties of the equation itself. For example, 
the Ermakov--Pinney equation \cite{ermakov, pinney} which in its 
simplest form is
\begin{gather}
w'' + \frac {K} {w ^ 3} = 0, \label{1.1}
\end{gather}
where $K$ is a constant.  In theoretical discussions the sign of  
the constant $K $ is immaterial and in fact it is often rescaled 
to unity. 

The general form of (\ref {1.1}), {\it videlicet}
\begin{gather*}
\ddot {\rho} +\omega ^ 2 (t)\rho = \frac {1} {\rho ^ 3} 
\end{gather*}
occurs in the study of the time-dependent linear oscillator,  be 
it the classical or the quantal problem, as the differential 
equation which determines the time-dependent rescaling of the 
space variable and the definition of `new time'. Some of the 
references for this are \cite{lewis67, lewis68}. 

Another origin of (\ref {1.1}) --- of particular interest in this 
work --- is as an integral of the third-order equation of maximal 
symmetry which in its elemental form is $y'''= 0$.

\section{Equations of maximal symmetry}

\begin{definition} We define a first integral $I$ for an
equation of maximal symmetry, $E=y^{(n)}=0$, as
$I=f\big(y,y',y',\ldots ,y^{(n-1)}\big)$, where 
\begin{gather*}
\left.\frac{{\d}I}{{\d}x}\right|_{E=0}=0 \quad \iff \quad
\left.\frac{{\d}f}{{\d}x}\right|_{E=0}=0.
\end{gather*} This means that, if $g\big(x,y,y',y'',\ldots ,y^{(n-1)}\big)$ is an
integrating factor, then 
\begin{gather*} 
\left.\frac{{\d}I}{{\d}x}\right|_{E=0}=\left.g 
E\left(x,y,y',\ldots ,y^{(n)}\right)\right|_{E=0}=0. 
\end{gather*}
\end{definition}

We start by considering the well-known third-order ordinary differential equation
of maximal symmetry 
\begin{gather} 
y'''=0\label{1} 
\end{gather}
 which has seven Lie point
symmetries. These are
\begin{gather}
G_{1} = \partial_{y},\qquad
G_{2} = x \partial_{y}, \qquad
G_{3} = x^{2} \partial_{y}, \qquad
G_{4} = y\partial_{y},\nonumber\\
G_{5} = \partial_{x},\qquad
G_{6}= x\partial_{x} + y\partial_{y},\qquad
G_{7} = x^2 \partial_{x} + 2xy\partial_{y}.\label{3aa}
\end{gather}
The algebra is $\{A_{1}\oplus_{s} sl(2,{\mathbb R})\}\oplus_s 
3A_{1}$. The autonomous integrating factors for \eqref{1} are 
$y''$ and~$y$. We list the symmetries and algebra when each of 
the integrals is treated as an equation and as a function.

When we multiply $y'''=0$ by the integrating factor $y''$, 
we obtain $y'' y'''=0$. Integration of this expression 
gives $\mbox{$\frac{1}{2}$} \left({y''}\right){}^{2}=k$, 
where $k$ is a constant of integration. This gives rise to three cases which we list as follows:
\begin{gather*}\arraycolsep=0em
\begin{array}{lll}
\multicolumn{1}{c}{y''=0} & &\multicolumn{1}{c}{y'' = k}\vspace{1mm}\\
G_1 = \partial_{y}, && G_1 = \partial_{y},\vspace{1mm}\\
G_2 = x\partial_{y}, && G_2 = x\partial_{y},\vspace{1mm}\\
G_3 = y \partial_{y}, && G_3 =\left(\dfrac{1}{2} x^2 k - y\right)\partial_{y},\vspace{1mm}\\
G_4 = \partial_{x},& & G_{4} = \partial_{x} + 2xk\partial_{y},\vspace{1mm}\\
G_5 = x\partial_{ x}, && G_5 = x \partial_{x} + x^2 k \partial_{y},\vspace{1mm}\\
G_6 = x^2\partial_{x}+ x y \partial_{y}, &\qquad\qquad\qquad & G_6 = x^2 \partial_{x} + 
\left(xy+\dfrac{1}{2} x^3 k\right)\partial_{y},\vspace{1mm}\\
G_7 = y\partial_{x},&& G_7 = \left(y-\dfrac{3}{2} x^2 k\right) \partial_{x} - x^3 k^2 \partial_{y}, \vspace{1mm}\\
G_8 = xy \partial_{x}+ y^2\partial_{y},& &G_8 = \left(x y - 
\dfrac{1}{2} x^3 k\right) \partial_{x} + \left(y^2 - \dfrac{1}{4} x^4 
k^2\right)\partial_{y},
\end{array}
\end{gather*}
and, when $y'' = k$ is treated as a function, we have 
\begin{gather*}
G_{1} = \partial_{y},\qquad
G_{2} = x \partial_{y},\qquad
G_{3} = \partial_{x},\qquad
G_{4} = x\partial_{x}+2y\partial_{y}.
\end{gather*}

\begin{remark}
When $y'' = k$ is treated as an equation, we have
two cases, that is, $y''=0$ and $y''=k$ for which the algebra is
$sl(3,{\mathbb R}): 2A_{1} \oplus_{s} \{sl(2,{\mathbb R})\oplus A_{1}\}\oplus 2
A_{1}$ \cite{mubarakzyanov631, mubarakzyanov632, mubarakzyanov633}. 
If $y''=k$ is treated as a function, the algebra is
$A_{4,9}^{1}: A_{2} \oplus_{s} 2 A_{1}$ \cite{mubarakzyanov631, mubarakzyanov632, mubarakzyanov633, paterasharpwinternitz}.
\end{remark}

If $y$ is used an the integrating factor, we obtain $y y''' = 0$.
Integration of this equation gives $y y'' - \frac{1}{2} {y'}{}^2 = k$ which can
be written as $\big(y^{1/2}\big)'' = {k}/{\big(y^{1/2}\big)^{3}}$ and is the
simplest form of the Ermakov--Pinney equation \cite{ermakov, pinney}.
As before we write down the point symmetries corresponding to the
three cases of the differential equation $u'' = k/u^{3}$, where $u
= y^{1/2}$. Program {\tt LIE}~\cite{head} gives the following:
\begin{gather*}\arraycolsep=0em
\begin{array}{lllll}
\multicolumn{1}{c}{u''=0}  &&\multicolumn{1}{c}{u'' = k/u^3} &&\multicolumn{1}{c}{u'' = k/u^3}\vspace{1mm}\\
G_1 = \partial_{u}, && G_1 = \partial_{x},&& G_{1} = \partial_{x},\vspace{1mm}\\
G_2 = x\partial_{u}, && G_2 = 2x\partial_{x}+u\partial_{u},&& G_2 = 2x\partial_{x}+u\partial_{u},\vspace{1mm}\\
G_3 = u \partial_{u}, && G_3 =x^{2}\partial_{x}+xu\partial_{u},&&G_3 =x^{2}\partial_{x}+xu\partial_{u},\vspace{1mm}\\
G_4 = \partial_{x},&&  &&\vspace{1mm}\\
G_5 = x\partial_{ x}, &&&&\vspace{1mm}\\
G_6 = x^2\partial_{x}+ x u \partial_{u}, &\qquad\qquad &&\qquad\qquad&\vspace{1mm}\\
G_7 = u\partial_{x},&& && \\
G_8 = xu \partial_{x}+ u^2\partial_{u}.&&&&
\end{array}
\end{gather*}
The transformation of $y y'' - \frac{1}{2} {y'}{}^2 = k$ to $u'' = k/u^{3}$
does not make a difference in terms of the symmetries as we just
have a point transformation in this case. The other obvious
integrating factors for \eqref{1} are $1$, $x$ and $\frac{1}{2} x^{2}$ which give
\begin{gather*}\arraycolsep=0em
\begin{array}{rcl}
1\cdot y'''= 0 &\quad\longrightarrow\quad & I_3 = y'',\vspace{1mm}\\
x\cdot y'''= 0 &\quad\longrightarrow\quad & I_2 = xy''-y',\vspace{1mm}\\
\dfrac{1}{2} x ^ 2\cdot  y'''= 0 &\quad\longrightarrow\quad & I_1 = \dfrac{1}{2} x ^ 2y''-xy'+y.
\end{array}
\end{gather*}
(Note that the numbering of the fundamental first integrals
follows the convention given in Flessas {\it et al} \cite{Flessas94, Flessas97}.)
\begin{itemize}
\itemsep=0pt
\item
 The
integration of equation \eqref{1}, which is a feature of the
calculation of the symmetries of all linear ordinary differential
equations of maximal symmetry \cite{mahomed90}, by means of 
an~integrating factor gives a variety of results depending upon the
integrating factor used.
\item The characteristic feature of the
Ermakov--Pinney equation is that it possesses the three-element
algebra of Lie point symmetries, $sl(2,{\mathbb R})$, which in itself is
characteristic of all scalar ordinary differential equations of
maximal symmetry.
\end{itemize}

The fourth-order ordinary differential equation $y^{\mbox{\scriptsize iv}}=0$ has
autonomous integrating factors~$y'$ and $y'''$. If we use $y'$ as an integrating factor in
the original equation and integrate, we obtain 
\begin{gather}
y'y'''-\frac{1}{2}(y''){}^2=k.\label{9a}
\end{gather}
Equation \eqref{9a} is a generalised Kummer--Schwartz equation 
for $k=0$ and for $k\neq 0$ a variation on the Ermakov--Pinney 
equation as it can be written in the form 
\begin{gather*}
\left((y')^{1/2}\right)''=\left(k\Big/\left((y')^{3/2}\right)\right).\nonumber
\end{gather*}
The three cases for
the integral in \eqref{9a} treated as an equation and as a function
give the following results:
\begin{gather*}\arraycolsep=0em
\begin{array}{lclcl}
\multicolumn{1}{c}{y'y'''-\dfrac{1}{2}({y''})^2=0}  &\qquad& 
\multicolumn{1}{c}{y'y'''-\dfrac{1}{2}(y'')^2=k} &\qquad& 
\multicolumn{1}{c}{y'y'''-\dfrac{1}{2}(y'')^2=k}\vspace{1mm}\\
G_1 = \partial_{x}, && G_1 = \partial_{x},&& G_{1} = \partial_{x},\vspace{1mm}\\
G_2 = x\partial_{x}, && G_2 = \partial_{y},&& G_{2} =  \partial_{y},\vspace{1mm}\\
G_3 = y \partial_{y}, && G_3 =x\partial_{x}+2y\partial_{y},&&G_{3} = x\partial_{x}+2y\partial_{y},\vspace{1mm}\\
G_4 = \partial_{y}.&& &&
\end{array}
\end{gather*}

The use of $y'''$ as an integrating factor gives $y'''=k$. If $k=0$, then we just have seven point symmetries 
as those of equation \eqref{1}.
The two remaining cases give
\begin{gather*}\arraycolsep=0em
\begin{array}{lcl}
\multicolumn{1}{c}{y'''=k}  &\qquad &\multicolumn{1}{c}{y'''=k}\vspace{1mm}\\
G_1 = \partial_{x}, && G_1 = \partial_{x}, \vspace{1mm}\\
G_2 = x\partial_{y}, && G_2 = \partial_{y},\vspace{1mm}\\
G_3 = \dfrac{1}{2}x^{2}\partial_{y}, && G_3 =x\partial_{y},\vspace{1mm}\\
G_4 = \partial_{y},&&G_4=x^{2}\partial_{y},\vspace{1mm}\\
G_5=x\partial_{x}+\dfrac{1}{2}{x^{3}}k\partial_{y}&&G_{5}=x\partial_{x}+3y\partial_{y},\vspace{1mm}\\
G_{6}=\left(y-\dfrac{1}{6}x^{3}k\right)\partial_{y},&&\vspace{1mm}\\
G_{7}=x^{2}\partial_{x}+\left(2xy+\dfrac{1}{6}{x^{4}}k\right)\partial_{y}. &&
\end{array}
\end{gather*}

Consider the fifth-order equation of maximal symmetry given by
\begin{gather}
y^{\mbox{\scriptsize v}}=0\label{2}
\end{gather}
with autonomous integrating factors $y$, $ y''$ and $y^{\mbox{\scriptsize iv}}$. 
If we multiply \eqref{2} by the first integrating factor and integrate, we obtain the integral
\begin{gather}
yy^{\mbox{\scriptsize iv}}-y' y'''+\frac{1}{2}( y'')^{2}=k.\label{3}
\end{gather}
We consider the three cases for \eqref{3} treated as an equation with $k=0$, $k\neq 0$ and as a function.
\begin{gather*}\arraycolsep=0em
\begin{array}{lclcl}
\multicolumn{1}{c}{yy^{\mbox{\scriptsize iv}}-y' y'''+\dfrac{1}{2}(y'')^{2}=0}  
&\qquad & \multicolumn{1}{c}{yy^{\mbox{\scriptsize iv}}-y' y'''+\dfrac{1}{2}(y'')^{2}=k} 
&\qquad &\multicolumn{1}{c}{yy^{\mbox{\scriptsize iv}}-y' y'''+\dfrac{1}{2} (y'')^{2}=k}\vspace{1mm}\\
G_1 = \partial_{x}, && G_1 = \partial_{x},&& G_{1} = \partial_{x},\vspace{1mm}\\
G_2 = x\partial_{x}, && G_2 = x\partial_{x}+2y\partial_{y},&& G_{2} = x\partial_{x}+2y\partial_{y},\vspace{1mm}\\
G_3 = y\partial_{y},&&G_3 = x^{2}\partial_{x}+4xy\partial_{y},&&G_3 = x^{2}\partial_{x}+4xy\partial_{y},\vspace{1mm}\\
G_4 = x^{2}\partial_{x}+4xy\partial_{y}. &&&&
\end{array}\hspace{-10mm}
\end{gather*}
\begin{remark}
For easier closure of the algebra in the first case $x\partial{x}$ can be written as $x\partial_{x}+2y\partial_{y}$.
\end{remark}

\begin{itemize}
\itemsep=0pt
\item
 We also observe that there is no difference when the
integral is treated as a function and as an equation. It is
important to note that, if $y$ is an integrating factor of
$y^{(n)}=0$, then the integral obtained using this integrating
factor always has the $sl(2,{\mathbb R})$ subalgebra. 
\item We further
observe that for the peculiar value of the constant, that is, $k=0$, there is the
splitting of the self-similarity symmetry into two homogeneity
symmetries.
\end{itemize}

The integrating factor $y''$ with \eqref{2} gives the
following results
\begin{gather*}\arraycolsep=0em
\begin{array}{lclcl}
\multicolumn{1}{c}{y''y^{\mbox{\scriptsize iv}}-\dfrac{1}{2}  (y''')^{2}=0}  
&\qquad & \multicolumn{1}{c}{y''y^{\mbox{\scriptsize iv}}-\dfrac{1}{2}  (y''')^{2}=k} 
&\qquad &
\multicolumn{1}{c}{y''y^{\mbox{\scriptsize iv}}-\dfrac{1}{2} (y''')^{2}=k}\vspace{1mm}\\
G_1 = \partial_{x}, && G_1 = \partial_{x},&& G_{1} = \partial_{x},\vspace{1mm}\\
G_2 = x\partial_{x}, && G_2 = \partial_{y},&& G_{2} = \partial_{y},\vspace{1mm}\\
G_3 = \partial_{y}, && G_3 =x\partial_{y},&&G_{3} =x \partial_{y},\vspace{1mm}\\
G_4 = x\partial_{y},&&G_4=x\partial_{x}+3y \partial_{y},&&G_4=x\partial_{x}+3y \partial_{y},\vspace{1mm}\\
G_5=y\partial_{y}.&&&&
\end{array}
\end{gather*}
If we use $y^{\mbox{\scriptsize iv}}$ as the integrating factor of \eqref{2} and integrate, we obtain
\begin{gather*}
y^{\mbox{\scriptsize iv}}=k. 
\end{gather*}
We delineate the three cases below:
\begin{gather*}\arraycolsep=0em
\begin{array}{lclcl}
\multicolumn{1}{c}{y^{\mbox{\scriptsize iv}}=0}  &\qquad &
\multicolumn{1}{c}{y^{\mbox{\scriptsize iv}}=k} &\qquad &
\multicolumn{1}{c}{y^{\mbox{\scriptsize iv}}=k}\vspace{1mm}\\
G_1 = \partial_{y}, && G_1 = \partial_{y}&& G_{1} = \partial_{y},\vspace{1mm}\\
G_2 = x\partial_{y}, && G_2 = x\partial_{y}&& G_{2} = x \partial_{y},\vspace{1mm}\\
G_3 = x^{2} \partial_{y}, && G_3 =\dfrac{1}{2} x^2 \partial_{y}&&G_{3} =x^{2}\partial_{y},\vspace{1mm}\\
G_4 = x^{3} \partial_{y}, &&  G_{4} = \dfrac{1}{6}x^{3}\partial_{y}&&G_{4}= x^{3}\partial_{y},\vspace{1mm}\\
G_5 = y\partial_{ y}, &&G_5 =  \partial_{x}&&G_{5}=\partial_{x},\vspace{1mm}\\
G_6 = \partial_{x}, &&G_6 =6x\partial_{x} + x^{4}k\partial_{y}&&G_{6}=x\partial_{x}+4y\partial_{y},\vspace{1mm}\\
G_7 = x\partial_{x},&& G_7 = \big(24y-x^{3}k\big) \partial_{y},&& \vspace{1mm}\\
G_8 = x^{2}\partial_{x}+ 3xy\partial_{y},&&G_8 = 24 x^2 \partial_{x} + \big(72xy+x^{5}k\big)\partial_{y}.&&
\end{array}
\end{gather*}
The differential equation
\begin{gather}
y^{\mbox{\scriptsize vi}}=0\label{4}
\end{gather}
has integrating factors $y'$, $y'''$ and $y^{\mbox{\scriptsize v}}$. If we use $y'$ as the integrating factor, we obtain 
\begin{gather*}
y'y^{\mbox{\scriptsize v}}-y''y^{\mbox{\scriptsize iv}}+\frac{1}{2}( y''')^{2}=k
\end{gather*}
which leads to the cases below.
\begin{gather*}\arraycolsep=0em
\begin{array}{lclcl}
\multicolumn{1}{c}{y'y^{\mbox{\scriptsize v}}-y''y^{\mbox{\scriptsize iv}}+
\dfrac{1}{2} (y''')^{2}=0}  &\qquad &
\multicolumn{1}{c}{y'y^{\mbox{\scriptsize v}}-y''y^{\mbox{\scriptsize iv}}+\dfrac{1}{2} (y''')^{2}=k}
 &\qquad &
\multicolumn{1}{c}{y'y^{\mbox{\scriptsize v}}-y''y^{\mbox{\scriptsize iv}}+\dfrac{1}{2}(y''')^{2}=k}\vspace{1mm}\\
G_1 = \partial_{y}, && G_1 = \partial_{y},&& G_{1} = \partial_{y},\vspace{1mm}\\
G_2 = y\partial_{y}, && G_2 = \partial_{x},&& G_{2} =  \partial_{x},\vspace{1mm}\\
G_3 = \partial_{x}, && G_3 =x \partial_{x}+3y\partial_{y},&&G_{3} =x\partial_{x}+3y\partial_{y},\vspace{1mm}\\
G_4 = x \partial_{x}.&& &&
\end{array}\hspace{-10mm}
\end{gather*}

The use of $y'''$ as the integrating factor for \eqref{4} leads to
\begin{gather*}
y'''y^{\mbox{\scriptsize v}}-\frac{1}{2}\big( y^{\mbox{\scriptsize iv}}\big)^2=k.
\end{gather*}

The three cases give the following results:
\begin{gather*}\arraycolsep=0em
\begin{array}{lclcl}
\multicolumn{1}{c}{y'''y^{\mbox{\scriptsize v}}-\dfrac{1}{2} \big(y^{\mbox{\scriptsize iv}}\big)^2=0}  &\qquad&
\multicolumn{1}{c}{y'''y^{\mbox{\scriptsize v}}-\dfrac{1}{2} \big(y^{\mbox{\scriptsize iv}}\big)^2=k} &\qquad &
\multicolumn{1}{c}{y'''y^{\mbox{\scriptsize v}}-\dfrac{1}{2} \big(y^{\mbox{\scriptsize iv}}\big)^2=k}\vspace{1mm}\\
G_1 = \partial_{y}, && G_1 = \partial_{y},&& G_{1} = \partial_{y},\vspace{1mm}\\
G_2 = \partial_{x}, && G_2 = \partial_{x},&& G_{2} =  \partial_{x},\vspace{1mm}\\
G_3=x\partial_{y}, &&G_3=x\partial_{y}, &&G_3=x\partial_{y},\vspace{1mm}\\
G_4=x^{2}\partial_{y},&&G_4=x^{2}\partial_{y}, &&G_4=x^{2}\partial_{y},\vspace{1mm}\\
G_5 = y\partial_{y}, &&G_5 = x\partial_{x}+ 4y\partial_{y}, &&G_5 = x\partial_{x}+ 4y\partial_{y},\vspace{1mm}\\
G_6 =x \partial_{x}.&&&&
\end{array}
\end{gather*}
If we use $y^{\mbox{\scriptsize v}}$ as an integrating factor, we obtain
\begin{gather*}
y^{\mbox{\scriptsize v}}=k.
\end{gather*}
We also have the three cases as mentioned above to obtain
\begin{gather*}\arraycolsep=0em
\begin{array}{lclcl}
\multicolumn{1}{c}{y^{\mbox{\scriptsize v}}=0}  &\qquad &
\multicolumn{1}{c}{y^{\mbox{\scriptsize v}}=k} &\qquad &
\multicolumn{1}{c}{y^{\mbox{\scriptsize v}}=k}\vspace{1mm}\\
G_1 = \partial_{y}, && G_1 = \partial_{y},&& G_{1} = \partial_{y},\vspace{1mm}\\
G_2 = y\partial_{y}, && G_2 = \partial_{x},&& G_{2} =  \partial_{x},\vspace{1mm}\\
G_3 = \partial_{x}, &&G_3 = x\partial_{y}, &&G_3 = x\partial_{y},\vspace{1mm}\\
G_4=x\partial_{y},&&G_4=x^{2}\partial_{y},&&G_4=x^{2}\partial_{y},\vspace{1mm}\\
G_5=x^{2}\partial_{y},&&G_5=x^{3}\partial_{y},&&G_5=x^{3}\partial_{y},\vspace{1mm}\\
G_6=x^{3}\partial_{y},&&G_6=x^{4}\partial_{y},&&G_6=x^{4}\partial_{y},\vspace{1mm}\\
G_7=x^{4}\partial_{y},&&G_7=x\partial_{x}+\dfrac{1}{24}kx^{5}\partial_{y},&&G_7=x\partial_{x}+5y\partial_{y},\vspace{1mm}\\
G_8=x\partial_{x},&&G_8=\left(y-\dfrac{1}{120}kx^{5}\right)\partial_{y},&&\vspace{1mm}\\
G_9=x^{2}\partial_{x}+4xy\partial_{y},&&G_9=x^{2}\partial_{x}+\left(4xy+\dfrac{1}{120}kx^{6}\right)\partial_{y}.&&
\end{array}
\end{gather*}

For the differential equation
$
y^{\mbox{\scriptsize vii}}=0
$
we have the integrating factors $y$, $y''$, $y^{\mbox{\scriptsize  iv}}$ 
and $y^{\mbox{\scriptsize vi}}$. The integrals corresponding to these integrating factors respectively are
\begin{gather}\arraycolsep=0em
\begin{array}{l}
yy^{\mbox{\scriptsize vi}}-y'y^{\mbox{\scriptsize v}}+y''y^{\mbox{\scriptsize iv}}-\dfrac{1}{2}(y''')^2=k,\vspace{1mm}\\
y''y^{\mbox{\scriptsize vi}}-y'''y^{v}+\dfrac{1}{2}\big(y^{\mbox{\scriptsize iv}}\big)^2=k,\vspace{1mm}\\
y^{\mbox{\scriptsize iv}}y^{\mbox{\scriptsize vi}}-\dfrac{1}{2}\big(y^{\mbox{\scriptsize v}}\big)^{2}=k,\vspace{1mm}\\
y^{\mbox{\scriptsize vi}}=k.
\end{array}
\end{gather}
If $y$ is used as the integrating factor, we have 
the integral $yy^{\mbox{\scriptsize vi}}-y'y^{\mbox{\scriptsize v}}+y''y^{\mbox{\scriptsize iv}}-\frac{1}{2}
(y''')^2=k$ 
which is treated as an equation for $k=0$, $k\neq 0$ and as a function. This gives the following results:
\begin{gather*}\arraycolsep=0em
\begin{array}{lclcl}
G_1 = \partial_{x}, &\qquad & G_1 = \partial_{x},&\qquad& G_{1} = \partial_{y},\vspace{1mm}\\
G_2 = x\partial_{x}, && G_2 = x\partial_{x}+3y\partial_{y},&& G_2 = x\partial_{x}+3y\partial_{y},\vspace{1mm}\\
G_3 = y\partial_{y}, &&G_3 = x^{2}\partial_{x}+6xy\partial_{y}, &&G_3 = x^{2}\partial_{x}+6xy\partial_{y},\vspace{1mm}\\
G_4=x^{2}\partial_{x}+6xy\partial_{y}.&&&&
\end{array}
\end{gather*}
The integral corresponding to the integrating factor $y''$ leads to the following cases:
\begin{gather*}\arraycolsep=0em
\begin{array}{lclcl}
\multicolumn{1}{c}{y''y^{\mbox{\scriptsize vi}}-y'''y^{\mbox{\scriptsize v}}+\dfrac{1}{2}
\big(y^{\mbox{\scriptsize iv}}\big)^2=0}  &\ \quad &
\multicolumn{1}{c}{y''y^{\mbox{\scriptsize vi}}-y'''y^{\mbox{\scriptsize v}}+\dfrac{1}{2}
\big(y^{\mbox{\scriptsize iv}}\big)^2=k} 
&\ \quad
&\multicolumn{1}{c}{y''y^{\mbox{\scriptsize vi}}-y'''y^{\mbox{\scriptsize v}}+\dfrac{1}{2}
\big(y^{\mbox{\scriptsize iv}}\big)^2= k}\vspace{1mm}\\
G_1 = \partial_{y}, && G_1 = \partial_{y},&& G_{1} = \partial_{y},\vspace{1mm}\\
G_2 = \partial_{x}, && G_2 = \partial_{x},&& G_{2} =  \partial_{x},\vspace{1mm}\\
G_3 = x\partial_{y}, &&G_3 = x\partial_{y}, &&G_3 = x\partial_{y},\vspace{1mm}\\
G_4=x\partial_{x},&&G_4=x\partial_{x}+4y\partial_{y},&&G_4=x\partial_{x}+4y\partial_{y},\vspace{1mm}\\
G_5=y\partial_{y}.&&&&
\end{array}
\end{gather*}
For the integrating factor $y^{\mbox{\scriptsize iv}}$ we have the cases:
\begin{gather*}\arraycolsep=0em
\begin{array}{lclcl}
\multicolumn{1}{c}{y^{\mbox{\scriptsize iv}}y^{\mbox{\scriptsize vi}}
-\dfrac{1}{2}\big(y^{\mbox{\scriptsize v}}\big)^{2}=0} 
 &\qquad &
\multicolumn{1}{c}{y^{\mbox{\scriptsize iv}}y^{\mbox{\scriptsize vi}}
-\dfrac{1}{2}\big(y^{\mbox{\scriptsize v}}\big)^{2}=k} &\qquad &
\multicolumn{1}{c}{y^{\mbox{\scriptsize iv}}y^{\mbox{\scriptsize vi}}
-\dfrac{1}{2}\big(y^{\mbox{\scriptsize v}}\big)^{2}=k}\vspace{1mm}\\
G_1 = \partial_{y}, && G_1 = \partial_{y},&& G_{1} = \partial_{y},\vspace{1mm}\\
G_2 = \partial_{x}, && G_2 = \partial_{x},&& G_{2} =  \partial_{x},\vspace{1mm}\\
G_3 = y\partial_{y}, &&G_3 = x\partial_{y}, &&G_3 = x\partial_{y},\vspace{1mm}\\
G_4 = x\partial_{y},&&G_4=x^{2}\partial_{y},&&G_4=x^{2}\partial_{y},\vspace{1mm}\\
G_5=x^{2}\partial_{y},&&G_5=x^{3}\partial_{y},&&G_5=x^{3}\partial_{y},\vspace{1mm}\\
G_6=x^{3}\partial_{y},&&G_6=x\partial_{x}+5y\partial_{y},&&G_6=x\partial_{x}+5y\partial_{y},\vspace{1mm}\\
G_7=x\partial_{x}.&&&&
\end{array}
\end{gather*}

The last of the four integrating factors $y^{\mbox{\scriptsize vi}}$
 leads to $y^{\mbox{\scriptsize vi}}=k$. We have for the three cases the following results:
\begin{gather*}\arraycolsep=0em
\begin{array}{lclcl}
\multicolumn{1}{c}{y^{\mbox{\scriptsize vi}}=0}  &\qquad &\multicolumn{1}{c}{y^{\mbox{\scriptsize vi}}=k} 
&\qquad &\multicolumn{1}{c}{y^{\mbox{\scriptsize vi}}=k}\vspace{1mm}\\
G_1 = \partial_{y}, && G_1 = \partial_{y},&& G_{1} = \partial_{y},\vspace{1mm}\\
G_2 = \partial_{x}, && G_2 = \partial_{x},&& G_{2} =  \partial_{x},\vspace{1mm}\\
G_3 = y\partial_{y}, &&G_3 = x\partial_{y}, &&G_3 = x\partial_{y},\vspace{1mm}\\
G_4=x\partial_{y},&&G_4=\dfrac{1}{2}x^{2}\partial_{y},&&G_4=x^{2}\partial_{y},\vspace{1mm}\\
G_5=x^{2}\partial_{y},&&G_5=\dfrac{1}{6}x^{3}\partial_{y},&&G_5=x^{3}\partial_{y},\vspace{1mm}\\
G_6=x^{3}\partial_{y},&&G_6=\dfrac{1}{24}x^{4}\partial_{y},&&G_6=x^{4}\partial_{y},\vspace{1mm}\\
G_7=x^{4}\partial_{y},&&G_7=\dfrac{1}{120}x^{5}\partial_{y},&&G_7=x^{5}\partial_{y},\vspace{1mm}\\
G_8=x^{5}\partial_{y},&&G_8=x\partial_{x}+\dfrac{1}{120}kx^6\partial_{y},&&G_8=x\partial_{x}+6y\partial_{y},\vspace{1mm}\\
G_9=x\partial_{x},&&G_9=\left(y-\dfrac{1}{720}kx^6\right)\partial_{y},&&\vspace{1mm}\\
G_{10}=x^{2}+5xy\partial_{y},&&G_{10}=x^{2}\partial_{x}+\left(5xy+\dfrac{1}{3600}k5x^{7}\right)\partial_{y}.&&
\end{array}
\end{gather*}

The differential equation
\begin{gather}
y^{\mbox{\scriptsize viii}}=0\label{6}
\end{gather}
has integrating factors $y'$, $y'''$, $y^{\mbox{\scriptsize v}}$ and $y^{\mbox{\scriptsize vii}}$.
If we use $y'$ in \eqref{6} and integrate the resulting equation, we obtain the integral
\begin{gather}
y'y^{\mbox{\scriptsize vii}}-y''y^{\mbox{\scriptsize vi}}+y'''y^{\mbox{\scriptsize v}}-
\frac{1}{2}\big(y^{\mbox{\scriptsize iv}}\big)^2=k.\label{7}
\end{gather}

The three cases of the integral in \eqref{7} 
being treated as an equation with $k=0$ and $k\neq 0$  and as a function are given
respectively below:
\begin{gather*}\arraycolsep=0em
\begin{array}{lclcl}
G_1 = \partial_{y} &\qquad & G_1 = \partial_{y}&\qquad & G_{1} = \partial_{y},\vspace{1mm}\\
G_2 = y\partial_{y}, && G_2 = \partial_{x},&& G_{2} =  \partial_{x},\vspace{1mm}\\
G_3 = \partial_{x}, &&G_3 = x\partial_{x}+4y\partial_{y}, &&G_3 = x\partial_{x}+4y\partial_{y},\vspace{1mm}\\
G_4=x\partial_{x}.&&&&
\end{array}
\end{gather*}

If $y'''$ is used as an integrating factor, we obtain
\begin{gather*}
y'''y^{\mbox{\scriptsize vii}}-y^{\mbox{\scriptsize iv}}y^{\mbox{\scriptsize vi}}
+\frac{1}{2}\big(y^{\mbox{\scriptsize v}}\big)^2=k
\end{gather*}
with the following respective cases:
\begin{gather*}\arraycolsep=0em
\begin{array}{lclcl}
G_1 = \partial_{y}, &\qquad& G_1 = \partial_{y},&\qquad& G_{1} = \partial_{y},\vspace{1mm}\\
G_2 = y\partial_{y}, && G_2 = \partial_{x},&& G_{2} =  \partial_{x},\vspace{1mm}\\
G_3 = \partial_{x}, &&G_3 = x\partial_{y}, &&G_3 = x\partial_{y},\vspace{1mm}\\
G_4=x\partial_{y},&&G_4=x^{2}\partial_{y},&&G_4=x^{2}\partial_{y},\vspace{1mm}\\
G_5=x^{2}\partial_{y}.&&&&
\end{array}
\end{gather*}

The use of $y^{\mbox{\scriptsize v}}$ as an integrating factor gives
\begin{gather}
y^{v}y^{\mbox{\scriptsize vii}}-\frac{1}{2}\big(y^{\mbox{\scriptsize vi}}\big){}^2=k.\label{9}
\end{gather}
Equation \eqref{9} is of the Ermakov--Pinney type. The three cases can be delineated as follows:
\begin{gather*}\arraycolsep=0em
\begin{array}{lclcl}
\multicolumn{1}{c}{y^{\mbox{\scriptsize v}}y^{\mbox{\scriptsize vii}}
-\frac{1}{2}\big(y^{\mbox{\scriptsize vi}}\big)^2=0}  &\qquad
&\multicolumn{1}{c}{y^{\mbox{\scriptsize v}}y^{\mbox{\scriptsize vii}}
-\frac{1}{2}\big(y^{\mbox{\scriptsize vi}}\big)^2=k} &\qquad
&\multicolumn{1}{c}{y^{\mbox{\scriptsize v}}y^{\mbox{\scriptsize vii}}
-\frac{1}{2}\big(y^{\mbox{\scriptsize vi}}\big)^2=k}\vspace{1mm}\\
G_1 = \partial_{y}, && G_1 = \partial_{y},&& G_{1} = \partial_{y},\vspace{1mm}\\
G_2 = y\partial_{y}, && G_2 = \partial_{x},&& G_{2} =  \partial_{x},\vspace{1mm}\\
G_3 = \partial_{x}, &&G_3 = x\partial_{y}, &&G_3 = x\partial_{y},\vspace{1mm}\\
G_4=x\partial_{y},&&G_4=x^{2}\partial_{y},&&G_4=x^{2}\partial_{y},\vspace{1mm}\\
G_5=x^{2}\partial_{y},&&G_5=x^{3}\partial_{y},&&G_5=x^{3}\partial_{y},\vspace{1mm}\\
G_6=x^{3}\partial_{y},&&G_6=x^{4}\partial_{y},&&G_6=x^{4}\partial_{y},\vspace{1mm}\\
G_7=x^{4}\partial_{y},&&G_7=x\partial_{x}+6y\partial_{y},&&G_7=x\partial_{x}+6y\partial_{y},\vspace{1mm}\\
G_8=x\partial_{x}.&&&&
\end{array}
\end{gather*}

If $y^{\mbox{\scriptsize vii}}$ is used as an integrating factor in \eqref{6}, 
we obtain $y^{\mbox{\scriptsize vii}}=k$ with the following symmetries for each of the three cases:
\begin{gather*}\arraycolsep=0em
\begin{array}{lclcl}
\multicolumn{1}{c}{y^{\mbox{\scriptsize vii}}=0}  &\qquad&
\multicolumn{1}{c}{y^{\mbox{\scriptsize vii}}=k} &\qquad &
\multicolumn{1}{c}{y^{\mbox{\scriptsize vii}}=k}\vspace{1mm}\\
G_1 = \partial_{y}, && G_1 = \partial_{y},&& G_{1} = \partial_{y},\vspace{1mm}\\
G_2 = y\partial_{y}, && G_2 = \partial_{x},&& G_{2} =  \partial_{x},\vspace{1mm}\\
G_3 = \partial_{x}, &&G_3 = x\partial_{y}, &&G_3 = x\partial_{y},\vspace{1mm}\\
G_4=x\partial_{y},&&G_4=\dfrac{1}{2}x^{2}\partial_{y},&&G_4=x^{2}\partial_{y},\vspace{1mm}\\
G_5=x^{2}\partial_{y},&&G_5=\dfrac{1}{6}x^{3}\partial_{y},&&G_5=x^{3}\partial_{y},\vspace{1mm}\\
G_6=x^{3}\partial_{y},&&G_6=\dfrac{1}{24} x^{4}\partial_{y},&&G_6=x^{4}\partial_{y},\vspace{1mm}\\
G_7=x^{4}\partial_{y},&&G_7=\dfrac{1}{120} x^{5} \partial_{y},&&G_7=x^{5}\partial_{y},\vspace{1mm}\\
G_8=x^{5}\partial_{y},&&G_8=\dfrac{1}{720}x^{6}\partial_{y},&&G_8=x^6\partial_{y},\vspace{1mm}\\
G_9=x^{6}\partial_{y},&&G_9=x\partial_{x}+\dfrac{1}{720}kx^{7}\partial_{y},&&G_{9}=x\partial_{x}+7y\partial_{y},\vspace{1mm}\\
G_{10}=x\partial_{x},&&G_{10}=\left(y-\dfrac{1}{5040}kx^{7}\right)\partial_{y},&&\vspace{1mm}\\
G_{11}=x^{2}\partial_{x}+6xy\partial_{y},&&G_{11}=x^{2}\partial_{x}+
\left(6xy+\dfrac{x^{8}}{5040}k\right)\partial_{y}.&&
\end{array}
\end{gather*}

\section{Relationship between fundamental integrals\\ and integrals obtained from integrating factors}
Consider the example of the third-order ordinary differential equation 
$
y'''=0
$
with the three fundamental integrals together with the appropriate 
associated point symmetries from the subalgebra $sl(2,{\mathbb R})$:
\begin{gather*}\arraycolsep=0em
\begin{array}{lcl}
G_7 = x^{2}\partial_{x}+2xy\partial_{y},&\qquad& I_1 = \dfrac{1}{2} x^{2} y''-xy' +y,\vspace{1mm}\\
G_6 = x\partial_{x}+y\partial_{y}, && I_2 = xy''-y',\vspace{1mm}\\
G_5 = \partial_{x}, && I_3=y''.\vspace{1mm}
\end{array}
\end{gather*}
The numbering of the symmetries follows that of the listing of Lie point 
symmetries in (\ref{3aa}) and the ordering of the integrals is in terms of their solution symmetries.
Then the autonomous integral associated with the integrating factor $y$ comes from the combination
\begin{gather*}
J=I_{1}I_{3}-\frac{1}{2} I_{2}^{2} =yy''-\frac{1}{2} {y'}{}^{2}.
\end{gather*}
\begin{proposition}
All the integrals obtained using $y$ as an integrating factor
always have the $sl(2,{\mathbb R})$ subalgebra whereas the fundamental
integrals only have one of the $sl(2,{\mathbb R})$ elements.
\end{proposition}
\begin{proof}
To prove the first proposition we consider the $sl(2,{\mathbb R})$
subalgebra $\Lambda_{1}=\partial_{x}$, $\Lambda_{2}=x\partial_{x}+y\partial_{y}$ and
$\Lambda_{3}= x^{2}\partial_{x}+2xy\partial_{y}$ and the fundamental integrals
$I_{1}$, $I_{2}$ and $I_{3}$ respectively. Then we have the
following:
\begin{gather*}\arraycolsep=0em
\begin{array}{lclcl}
\Lambda_{1}I_{1} = I_{2}, &\qquad& \Lambda_{2}I_{1}= I_{1},&\qquad& \Lambda_{3}I_{1}= 0,\vspace{1mm}\\
\Lambda_{1}I_{2}= I_{3}, && \Lambda_{2}I_{2}= 0,&& \Lambda_{3}I_{2}= -2I_{1},\vspace{1mm}\\
\Lambda_{1}I_{3}= 0, && \Lambda_{2}I_{3}= -I_{3},&&\Lambda_{3}I_{3}= -2I_{2}.
\end{array}
\end{gather*}

We also observe that $\Lambda_{i}J =0$ for $i=1,2,3$. In fact it is easy 
to show that $\Lambda_{i}J =\epsilon_{ijk}I_{j}I_{k}$. This is shown below as follows:
\begin{gather*}
\Lambda_{1}J = I_{2}I_{3}-I_{2}I_{3}=0,\\
\Lambda_{2}J = I_{1}I_{3}-I_{1}I_{3}=0,\\
\Lambda_{3}J = -2I_{2}I_{1}+2I_{1}I_{2}=0.\tag*{\qed}
\end{gather*}\renewcommand{\qed}{}
\end{proof}
In general we have
\begin{gather}
I_{ni}=\sum_{j=0}^{n-i-1}\frac{(-1)^{j}x^{(n-j-i-1)}}{(n-j-i-1)!}y^{(n-j-1)}, \qquad i=0,1,\ldots,n-1,
\end{gather}
so that for $n=3$, $I_{30}=I_{1}$, $I_{31}=I_{2}$ and $I_{32}=I_{3}$.
The symmetries $\Lambda_{1}, \Lambda_{2}$ and $\Lambda_{3}$ operating on the fundamental integrals then yield
in general
\begin{gather*}
\Lambda_{1}I_{ni} = I_{n,i+1},\\
\Lambda_{2}I_{ni}= (1-i)I_{ni},\\
\Lambda_{3}I_{ni}= -(n+i-3)(n-i)I_{n,i-1},\\
I_{nn}=0.
\end{gather*}
If we take for example $\Lambda_{3}I_{3i}= -i(3-i)I_{3,i-1}$ with $n=3$ and $i=0,1,2$, we obtain
\begin{gather*}
\Lambda_{3}I_{30}=0,\\
\Lambda_{3}I_{31}=-2I_{30},\\
\Lambda_{3}I_{32}=-2I_{31},
\end{gather*}
where as above $I_{30}=I_{1}$, $I_{31}=I_{2}$ and $I_{32}=I_{3}$.

\begin{proposition}[\cite{moyoleach}]
If we take the equation of maximal symmetry $y^{(n)}=0$, the
$sl(2,{\mathbb R})$ subalgebra maps back to itself and is preserved.
\end{proposition}

\begin{proposition}
For the fifth-order equation $y^{\mbox{\rm \scriptsize v}}=0$
 the autonomous integral emanating from the integrating factor $y$
  can be obtained from $J=I_{0}I_{4}-I_{1}I_{3}+\frac{1}{2} I_{2}^{2}$, where
\begin{gather*}
I_{0}= \frac{1}{24}x^{4}y^{\mbox{\rm \scriptsize iv}}-\frac{1}{6}x^{3}y'''+
\frac{1}{2} x^{2} y''-xy' +y,\\
I_{1}= \frac{1}{6}x^{3}y^{\mbox{\rm \scriptsize iv}}-\frac{1}{2} x^{2} y'''+xy'' -y',\\
I_{2}= \frac{1}{2} x^{2}y^{\mbox{\rm \scriptsize iv}}- x y'''+y'',\\
I_{3}= x y^{\mbox{\rm \scriptsize iv}}- y''',\\
I_{4}=y^{\mbox{\rm \scriptsize iv}}.
\end{gather*}
\end{proposition}

\begin{proposition}
The fourth-order equation also has an autonomous integral $J$
 defined as $J=I_{1}I_{3}-\frac{1}{2} I_{2}^{2}$, where
\begin{gather*}
I_{1}= \frac{1}{2} x^{2}y'''- x y''+y',\\ I_{2}= x y'''- y'',\\
I_{3}=y'''.
\end{gather*}
\end{proposition}

\begin{proposition}
It can be shown that the differential equation $y^{\mbox{\rm \scriptsize vi}}=0$ 
also has the autonomous integral $J$ which is defined as
$J=I_{0}I_{6}-I_{1}I_{5}+I_{2}I_{4}-\frac{1}{2} I_{3}^{2}$ with the $I_{i}$ $(i=0,6)$ being redefined appropriately.
\end{proposition}

\section{Conclusion}
If $y^{(n)}=f\big(x,y,y',\ldots ,y^{n-1}\big)$ is an $n$th-order ordinary 
differential equation and $g\big(x,y,y',\ldots ,$ $y^{n-1)}\big)$ 
$=k$ is an integral, the integral obtained by multiplying the 
equation by the integrating factor and integrating once possesses 
certain symmetries when treated as a function, an equation for 
the general constant and a configurational invariant (k=0). It is 
important to note that, if $y$ is an integrating factor of 
$y^{(n)}=0$, then the integral obtained using this integrating 
factor always has the $sl(2,{\mathbb R})$ subalgebra whereas the 
fundamental integrals only have one of the $sl(2,{\mathbb R})$ elements. We 
further observe that for the peculiar value of the constant, 
$k=0$, there is the splitting of the self-similarity symmetry 
into two homogeneity symmetries. The third-order ordinary 
differential equation is actually special and leads to the 
Ermakov--Pinney type equation. The fourth-order ordinary 
differential equation $y^{\mbox{\scriptsize iv}}=0$ has $y'$ as one of its 
autonomous integrating factors which leads together with the the 
original equation upon integration to the generalised 
Kummer--Schwartz equation. An extension to other types of 
equations will be completed in a separate contribution. The 
question of what Lie point symmetries of an ordinary differential 
equation are also shared by all its first integrals will form the 
basis for the next contribution.
                                                                                                                                   
\subsection*{Acknowledgments}
SM thanks the National Research Foundation of South Africa and the Durban Institute of Technology for their support.
PGLL thanks the University of KwaZulu-Natal for its continuing support.

\LastPageEnding
\end{document}